\documentclass[aps,pra,twocolumn]{revtex4}
\usepackage{graphicx}
\usepackage{bm}
\usepackage{color}
\usepackage{wasysym}

\usepackage[applemac]{inputenc}

\begin{document}
\preprint{}
\title{Detector self-tomography}
\author{Ra\'ul C\'onsul and Alfredo Luis}
\email{alluis@fis.ucm.es}
\homepage{http://www.ucm.es/info/gioq}
\affiliation{Departamento de \'{O}ptica, Facultad de Ciencias
F\'{\i}sicas, Universidad Complutense, 28040 Madrid, Spain}
\date{\today}

\begin{abstract}
We  present an intuitive model of detector self-tomography. Two identical realisations of the detector are illuminated by an entangled state that connects the joint statistics in a way in which each detector sees the other as a kind of mirror reflection.  A suitable analysis of the statistics reveals the possibility of fully characterizing the detector. We apply this idea to Bell-type experiments revealing their nonclassical nature. 
\end{abstract}

\maketitle

Quantum phenomena are always revealed by phenomena whose statistics cannot be accounted for in classical physics. According to Born's rule, the probability of each measurement of the observable $A$ with outcome $a$ is determined in a very symmetrical way  by the system state $| \psi \rangle$ and the measurement states $| a \rangle$.  This is, $p (a | \psi ) = | \langle a | \psi \rangle |^2$, where typically  $| a \rangle$ are the eigenvectors of the measured observable $A$. This is after all a quite interesting feature of the quantum theory. For example, in quantum optics the scalar product $\langle a | \psi \rangle$ reveals that detector states $|a \rangle$ must be formally described in terms of light states, in spite of being all them made of matter. So we can ascribe to detector states quantum-light properties in exactly the same way we do to light states. Similar reasoning holds if the light beam is mixed $\rho$ and the detector is described by a positive operator-valued measure (POVM) $\Delta (a)$ so that $p (a | \rho ) = \mathrm{tr} \left [ \rho \Delta (a) \right ]$.

The symmetry of Born's rule raises the question of whether the quantum paradoxical results might be ascribed to the measurement states as well as to system states \cite{LA17}. For this reason, it is interesting to study the characteristics of detectors from a fully quantum point of view. There are several ways to do this, that started from the idea of detector tomography  \cite{LS99a,LS99b,LS99c,JF01,ZDCJEPW12,BKSSV17,NJ19}. The one proposed in this work is an strategy of self-tomography in which a detector observes itself, as it were in front of a mirror, as a kind of self-calibration \cite{DML04,ZZZ20}. To achieve this we illuminate two identical realizations of the detector with an entangled state. Roughly speaking, because of quantum state reduction, the measurement performed by one of the detectors collapses the state of light illuminating the other one on the very same state associated to the measurement outcome \cite{LS98}. Thus one detector is illuminated by the internal state of the other one, so to speak. 

The objective of this work is to derive the statistics from this double measurement as a method of detector self-tomography alternative to the already known protocols of detector tomography \cite{LS99a, LS99b, LS99c}. We examine the possibility of extracting from the joint statistics the relevant information to characterise completely the detector. We then apply this idea to Bell-type experiments revealing their nonclassical nature. 

\bigskip

For the sake of simplicity, the physical system will be as simple as possible: this is a spin 1/2, or equivalently, the polarization of a photon. So our basic system will be two-dimensional, that is, a qubit. We include the possibility of mixed system states $ \rho $ 
\begin{equation}
\label{rho}
\rho = \frac{1} {2} \left (\sigma_0 + \boldsymbol{s} \cdot \boldsymbol{\sigma} \right),
\end {equation}
where $ \boldsymbol{\sigma} $ are the three Pauli matrices, $ \sigma_0 $ is the identity, and $ \boldsymbol{s} $ is a real three-dimensional vector with $ | \boldsymbol{ s} | \leq 1 $.

\bigskip

Measuring means establishing a correspondence between states $ \rho $ and probability distributions $ p (a) $ which are the statistics of some measured observable $A$ in the state $ \rho $. We will consider only two possible outcomes, which we will denote as $ a = \pm 1 $, or sometimes simply as $ a = \pm $. In the most general case the correspondence  $\rho \rightarrow p (a)$ is of the form
\begin{equation}
\label{est}
p (a) =  \mathrm{tr} \left [ \Delta(a) \rho \right ]  ,
\end{equation}
where $\Delta (a)$ is a POVM. In our dichotomic two-dimensional scenario the most general POVM for a generic observable $A$ is of the form, up to a trivial factor proportional to the identity,
\begin{equation}
\label{Delta}
\Delta ( a ) =  \frac{1}{2} \left  ( \sigma_0  + a \boldsymbol{S} \cdot \boldsymbol{\sigma} \right ) ,
\end{equation}
where $\boldsymbol{S}$ is a real three-dimensional vector with $ | \boldsymbol{S}  | \leq 1 $ that  completely characterizes the measurement up to a sign, since $\pm \boldsymbol{S}$ describe the same measurement. Using the properties  of Pauli matrices the statistics (\ref{est}) becomes
\begin{equation}
\label{sAs}
p (a) =  \mathrm{tr} \left [ \Delta(a) \rho \right ]  =  \frac{1}{2} \left  ( 1  + a \boldsymbol{s} \cdot \boldsymbol{S} \right ) .
\end{equation}
This highlights the symmetry of the Born rule (\ref{est}) between the state of the system $ \rho $ and the  detector $ \Delta (a) $. 

\bigskip

The process of self-tomography of the POVM $\Delta (a)$ is illustrated in Fig. 1. The source produces a two-mode entangled state in three possible versions
\begin{equation}
\label{ent}
| \psi \rangle_b = \frac{1}{\sqrt{2}} \left ( | + \rangle_{1,b} \otimes | + \rangle_{2,b} + | - \rangle_{1,b} \otimes | - \rangle_{2,b} \right ) ,
\end{equation}
where  $| \pm  \rangle_{j,b}$ are eigenstates of the Pauli matrix $\sigma_{j,b}$, where $j=1,2$ denotes the field mode and $b=x,y,z$ the corresponding Pauli matrix. Each mode  $j=1,2$ describes the light beam impinging on each detector. In one of the modes, say mode 1, we may insert a transparent plate altering the polarization state of the photon via a unitary transformation $U$, producing a rotation of the Stokes vector $\boldsymbol{s}$ implemented by an $3 \times 3$ orthogonal matrix $R$ 
\begin{equation}
\label{Rrho}
U \rho U^\dagger= \frac{1} {2} \left ( \sigma_0 +  \boldsymbol{s} R^t \boldsymbol{\sigma} \right ),
\end {equation}
where the superscript $t$ denotes matrix transposition. This is to say that the polarization state of the light modes illuminating each detector may be different.

\begin{figure}[htbp]
\centering
\includegraphics[width=6cm]{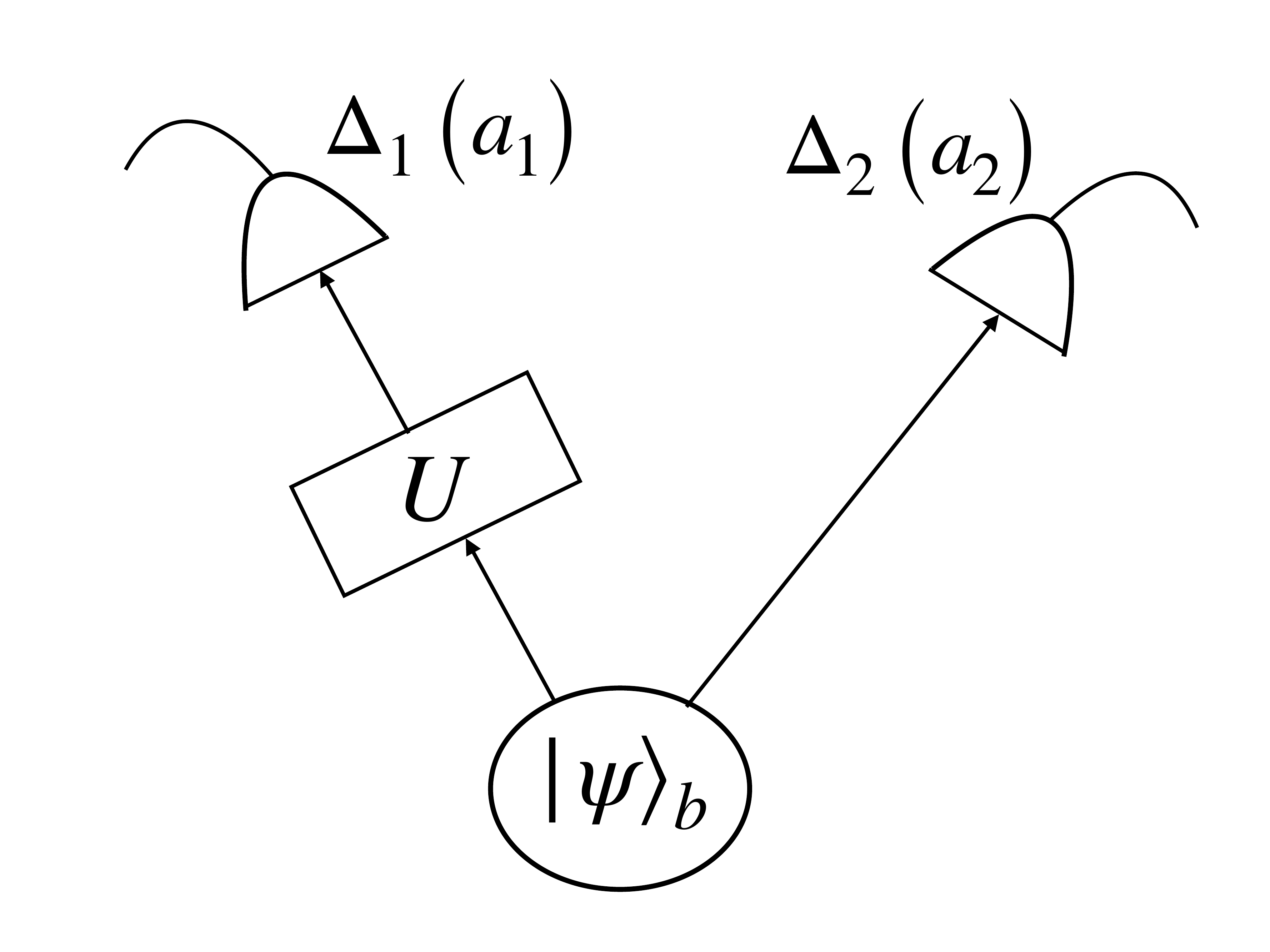}
\caption{Scheme of the self-tomography process.}
\label{fig:false-color}
\end{figure}

\bigskip
  
The joint statistics is 
\begin{equation}
\label{js}
p_{b,R}(a_1,a_2) =  {}_b \langle \psi | U^\dagger \Delta_1 (a_1) \otimes \Delta_2 (a_2 ) U |\psi \rangle_b ,
\end{equation}
where $\Delta_j (a_j)$ are two identical realizations of the same POVM $\Delta (a)$, leading to 
\begin{equation}
\label{est2}
p_{b,R} (a_1,a_2) = \frac{1}{4} \left ( 1 + a_1 a_2 \boldsymbol{S} R \boldsymbol{S}^\ast_b \right )  ,
\end{equation}
where $ \boldsymbol{S}^\ast_{b}$ are defined in terms of the components of $\boldsymbol{S} = (S_1, S_2, S_3 )  $ as
\begin{eqnarray}
& \boldsymbol{S}^\ast_x = (S_1, S_2, -S_3 )   , \quad \boldsymbol{S}^\ast_y = (-S_1, S_2, S_3 )  ,& \nonumber \\ &  \boldsymbol{S}^\ast_z = (S_1, -S_2, S_3 )   . &
\end{eqnarray}
This is that  $\boldsymbol{S}^\ast_{b}$ is the reflection of $\boldsymbol{S}$ in a coordinate plane. This relation (\ref{est2}) has essentially the same structure in Eq. (\ref{sAs}) where the system-state vector $\boldsymbol{s}$ defined by $\rho$ in Eq. (\ref{rho}) is replaced by a detector-state vector $\boldsymbol{S}$ defined by $\Delta (a)$ in Eq. (\ref{Delta}). This is the sense in which each detector sees the other.     

\bigskip

We can appreciate in Eq. (\ref{est2}) that the whole statistics is determined just from one result, say $a_1=a_2 =1$, since  
\begin{eqnarray}
& p_{b,R} (-1,-1) = p_{b,R} (1,1) , \\
& p_{b,R} (1,-1) = p_{b,R} (-1,1) = 1/2- p_{b,R} (1,1) . & \nonumber
\end{eqnarray}
We will use this fact to resume all statistics by just $ p_{b,R} (1,1)$, alleviating the notation writing $ p_{b,R} (1,1)$ simply as $p_{b,R} $.

\bigskip

Our objective here is self-tomography in the sense of a complete determination of $\boldsymbol{S}$ after the statistics  $p_{b,R}$. To this end we can consider the three bases $b=x,y,z$ and two choices for $R$, namely,
\begin{equation}
R_0 = \pmatrix{1 & 0 & 0 \cr 0 & 1 & 0 \cr 0 & 0 & 1} , \quad 
R_1 = \pmatrix{0 & 1 & 0 \cr 0 & 0 & 1 \cr 1 & 0 & 0} ,
\end{equation}
to give 
\begin{eqnarray}
\label{p0}
& 4 p_{x,0} = 1+ S_1^2 + S_2^2  - S_3^2 , & \nonumber \\
 &4 p_{y,0} = 1- S_1^2 + S_2^2   + S_3^2 , & \\
 &4 p_{z,0} = 1+S_1^2 - S_2^2   +S_3^2 , & \nonumber
\end{eqnarray}
and 
\begin{eqnarray}
\label{p1}
& 4 p_{x,1} = 1+ S_1 S_2  - S_2 S_3 + S_1 S_3 , & \nonumber \\
 &4 p_{y,1} = 1 + S_1 S_2  + S_2 S_3 - S_1 S_3, & \\
 &4 p_{z,1} = 1 - S_1 S_2  + S_2 S_3 + S_1 S_3 , & \nonumber
\end{eqnarray}
respectively. It is clear that these six equations fully determinate  $\boldsymbol{S}$ up to an irrelevant global sign, since $\boldsymbol{S}$ and $-\boldsymbol{S}$ are the same observable. 

\bigskip

More specifically, at least one of the components of $\boldsymbol{S}$ is different from zero, say $S_3$ without loss of generality. The modulus of $S_3$ can be obtained from the $p_{b,0}$ results in Eq. (\ref{p0}) simply as 
\begin{equation}
\label{sol1}
S_3 = \sqrt{2  p_{y,0} +  2p_{z,0} - 1} .
\end{equation}
Then from  Eq. (\ref{p1})  we can get the other components as 
\begin{equation}
\label{sol2}
S_1 = \frac{2  p_{x,1} +  2p_{z,1} - 1}{S_3}, \quad S_2 = \frac{2  p_{y,1} +  2p_{z,1} - 1}{S_3 }.
\end{equation}
This completes the proof that this scheme allows the complete determination of the detector POVM.  

\bigskip

So far we have assumed that the two detectors are identical, but it is worth analyzing the case when the POVM for one realization is different from the POVM of the other one. This can be easily taken into account in our scheme by allowing that each POVM $\Delta _j (a_j)$ is described by a different vector $\boldsymbol{S}_j$ so that the joint statistics (\ref{js}) becomes 
\begin{equation}
p_{b,R} (a_1,a_2) = \frac{1}{4} \left ( 1 + a_1 a_2 \boldsymbol{S}_1 R \boldsymbol{S}^\ast_{2,b}\right )  .
\end{equation}
From this point two routes might be followed. We may generalize the above approach by including new choices for $R$ in order to obtain enough number of equations to determine both vectors $\boldsymbol{S}_j$ from the recorded statistics. We have not pursued this possibility here since it would take us away from the central objective of this work. On the other hand, we may consider the differences in $\boldsymbol{S}_j$ as having a random origin, say $\boldsymbol{S}_j = \boldsymbol{S}+ \delta \boldsymbol{S}_j$, where $\delta \boldsymbol{S}$ are random small variations from one detector to another considered as a samples within a large series of otherwise identically prepared detectors. The proper arena to analyze this situation is an error analysis studying the robustness of the inferred $\boldsymbol{S}$ in Eqs. (\ref{sol1}) and (\ref{sol2}) against the fluctuations   $\delta \boldsymbol{S}$. A simple rough analysis shows that the uncertainty in the reconstructed $\boldsymbol{S}$  are of the same order of $\delta \boldsymbol{S}$, as it could be expected given the simple algebraic relations relating these quantities.

\bigskip

The scheme presented above may be generalized to other scenarios {\em mutatis mutandis}. As a simple illustration let us consider on/off photon detectors, such as avalanche photodiodes, that detect the presence of at least one photon. They can be used as the building block of photon-number-resolving detectors via multiplexing \cite{SVA12,JB19,PLFM20}. Let us consider the click (+) and no click (-) elements of the POVM as
\begin{equation}
    \Delta (-) = \left ( 1 - p_d \right ) \sum_{n=0}^\infty \left ( 1 - \eta \right )^n | n\rangle \langle n | , \quad \Delta (+) = I - \Delta (-) ,
\end{equation}
 where $\eta$ is the quantum efficiency, $p_d$ the dark-count probability, $I$ is the identity, and $|n\rangle$ are the photon number states \cite{JB19}. A suitable counterpart of the entangled state (\ref{ent}) in this context might be the two-mode squeezed vacuum state
 \begin{equation}
     |\xi \rangle = \sqrt{1 -|\xi |^2} \sum_{n=0}^\infty \xi^n | n \rangle_1 \otimes | n \rangle_2 , 
 \end{equation}
where $\xi$ is a parameter directly connected to the mean number of photons of this state $\bar{n} = 2 |\xi |^2 /(1-|\xi |^2)$, so that we have maximal entanglement as $\bar{n}$ tends to infinity. When this state illuminates two realizations of the on/off detector, we have 
\begin{equation}
p(j,k) =   \langle \xi |\Delta_1 (j) \otimes \Delta_2 (k ) |\xi \rangle,
\end{equation}
for $j,k=\pm$ that leads to 
\begin{equation}
\label{mm}
p(-,-) = \frac{\left ( 1 - p_d \right )^2}{1 +\bar{n} \eta \left (1-\eta/2 \right )} ,
\end{equation}
and 
\begin{equation}
\label{mM}
p(+,-) = p(-,+)= \frac{1 - p_d}{1 +\bar{n}\eta/2} - p(-,-) ,
\end{equation}
with $p(+,+) =1-2 p(+,-) - p(-,-)$. Then it is clear that the statistics $p(j,k)$ provides enough information to obtain the two parameters $p_d$ and $\eta$ of the detector after some simple algebra in Eqs. (\ref{mm}) and (\ref{mM}), more simple as $\bar{n} \rightarrow \infty$. 

\bigskip

We can generalize the above procedure to obtain the self-tomography of a generalized detector scheme represented by a POVM $\tilde{\Delta} (x,y)$ providing information about two incompatible dichotomic observables $X,Y$, instead of a single observable $A$. Since the observables $X,Y$ are incompatible the information provided is necessarily fuzzy, although it can be still complete, in the sense that we may recover from the statistics $\tilde{p} (x,y)$ complete exact information about the statistics of both  $X$ and $Y$. More specifically we have that the most general POVM conveying a fuzzy but complete observation of $X,Y$ can be expressed as \cite{HA14}
\begin{equation}
\label{tAB}
 \tilde{\Delta} (x,y) = \frac{1}{4} \left  [ \sigma_0 +  \tilde{\boldsymbol{S}} (x,y) \cdot  \boldsymbol{\sigma}  \right ] ,
\end{equation}
with $ x_, y =\pm 1$, and
\begin{equation}
\label{tS}
\tilde{\boldsymbol{S}} (x,y) = x \gamma_X \boldsymbol{S}_X + y \gamma_Y \boldsymbol{S}_Y+ x y \gamma_{XY} \boldsymbol{S}_{XY},  
\end{equation}
where all $\boldsymbol{S}_{X,Y,XY}$ are real, unit, three-dimensional vectors, and $\gamma_{X,Y, XY}$ are real nonnegative factors expressing the accuracy of the joint measurement. In any case, for all the outcomes $x,y$ we have always that $| \tilde{\boldsymbol{S}} (x,y)| \leq 1$,  so that $\tilde{\Delta}^\dagger (x,y)= \tilde{\Delta}  (x,y)$ and $\tilde{\Delta} (x,y) >0$. The vectors $\boldsymbol{S}_{X,Y}$ represent the observables $X$ and $Y$, respectively, as in Eq. (\ref{Delta}), while $\boldsymbol{S}_{XY}$ contains their correlations.

\bigskip

The very same procedure presented above for $\Delta (a)$ allows the self-tomography of the POVM  $\tilde{\Delta} \left ( x, y \right )$ replacing throughout the protocol $\boldsymbol{S}$ by $\tilde{\boldsymbol{S}} (x,y)$ to get
\begin{equation}
\label{tpbR}
\tilde{p}_{b,R} (x_1,y_1,x_2,y_2) = \frac{1}{16} \left [ 1 +  \tilde{\boldsymbol{S}} (x_1,y_1) R \tilde{\boldsymbol{S}}^\ast_b (x_2,y_2) \right ]  .
\end{equation}

\bigskip

Therefore, we can follow the same procedure of the preceding section consecutively for each one of the four vectors $\tilde{\boldsymbol{S}} (j,k)$, $j=\pm 1$, $k= \pm 1$, that become fully determined by the Eqs. (\ref{sol1}) and (\ref{sol2}) replacing $\boldsymbol{S}$ by $\tilde{\boldsymbol{S}} (j,k)$, $p_{b,0}$ and $p_{b,1}$ by
$p_{b,0} (j,k,j,k)$ and $p_{b,1} (j,k,j,k)$, and finally the factor 2 by a factor 8.

\bigskip

We can go further to retrieve the vectors $\boldsymbol{S}_X$, $\boldsymbol{S}_Y$, $\boldsymbol{S}_{XY}$ and the factors $\gamma_X$, $\gamma_Y$,$\gamma_{XY}$ as functions of $\tilde{\boldsymbol{S}} (j,k)$. After Eq. (\ref{tS}) and taking into account that the vectors $\boldsymbol{S}_X$, $\boldsymbol{S}_Y$, and $\boldsymbol{S}_{XY}$ are unit-modulus and all the $\gamma$ nonnegative, we have 
\begin{eqnarray}
&\boldsymbol{S}_X = \frac{\tilde{\boldsymbol{S}} (1,1)+\tilde{\boldsymbol{S}} (1,-1)}{|\tilde{\boldsymbol{S}} (1,1)+\tilde{\boldsymbol{S}} (1,-1)|}, \quad  \gamma_X = \frac{|\tilde{\boldsymbol{S}} (1,1)+\tilde{\boldsymbol{S}} (1,-1)|}{2}, & \nonumber \\ 
&\boldsymbol{S}_Y = \frac{\tilde{\boldsymbol{S}} (1,1)+\tilde{\boldsymbol{S}} (-1,1)}{|\tilde{\boldsymbol{S}} (1,1)+\tilde{\boldsymbol{S}} (-1,1)|}, \quad  \gamma_Y = \frac{|\tilde{\boldsymbol{S}} (1,1)+\tilde{\boldsymbol{S}} (-1,1)|}{2},  & \\ 
&\boldsymbol{S}_{XY} = \frac{\tilde{\boldsymbol{S}} (1,1)+\tilde{\boldsymbol{S}} (-1,-1)}{|\tilde{\boldsymbol{S}} (1,1)+\tilde{\boldsymbol{S}} (-1,-1)|}, \quad  \gamma_{XY} = \frac{|\tilde{\boldsymbol{S}} (1,1)+\tilde{\boldsymbol{S}} (-1,-1)|}{2}. & \nonumber
\end{eqnarray}
This proofs the self-tomography of $\tilde{\Delta} (x,y)$.

\bigskip

We find quite remarkable the parallelism of the scheme in Fig. 1 with a Bell-type measurement. In both cases we have two parties in modes 1 and 2 that share a maximally entangled state and perform local measurements on the corresponding modes, these are the measurements of $X,Y$ in mode 2 and the measurements of $U^\dagger XU, U^\dagger YU$ in mode 1. The key of Bell tests is to obtain two-mode, pairwise, joint statistics that taken together are incompatible with classical causal reasoning. To this end the two parties must perform alternatively the measurement of incompatible observables, such as the ones $X$ and $Y$ just considered above. 

In a recent work we have proposed  Bell-type experiments where all the exact statistics involved in the Bell test are obtained from a noisy joint measurement of all the observables in a single experimental arrangement \cite{MAL20,Muynck}. This is exactly the statistics in Eq. (\ref{tpbR}). Absolutely all the weird peculiarities of quantum correlations must be then contained in the observed noisy joint statistics $\tilde{p}_{b,R} (x_1,y_1,x_2,y_2)$. 

A convenient way to extract such nonclassical features is to apply to $\tilde{p}_{b,R} (x_1,y_1,x_2,y_2)$ an inversion procedure providing the exact statistics of all observables involved in the Bell tests. Roughly speaking, this consists on removing the extra fuzziness implied by the simultaneous observation of incompatible observables. This inversion lead us from  $\tilde{p}_{b,R} (x_1,y_1,x_2,y_2)$ to a new distribution  $p_{b,R} (x_1,y_1,x_2,y_2)$ whose marginals are the exact marginals for the four observables $X,Y$ in mode 2, and $U^\dagger XU, U^\dagger YU$ in mode 1, as well as their two-party, pairwise combinations. Formally, this inversion procedure carried out in detail in Ref. \cite{MAL20}, can be expressed as  
\begin{equation}
\tilde{\Delta} \left ( x, y \right )  \rightarrow \Delta \left ( x, y \right ) , 
\end{equation}
being 
\begin{equation}
\label{tAB}
\Delta (x,y) = \frac{1}{4} \left  [ \sigma_0 +  \boldsymbol{S} (x,y) \cdot  \boldsymbol{\sigma}  \right ]
,
\end{equation}
and
\begin{equation}
\label{inv}
 \boldsymbol{S} (x,y) = x \boldsymbol{S}_X + y \boldsymbol{S}_Y+ x y \frac{\gamma_{XY}}{\gamma_X \gamma_Y} \boldsymbol{S}_{XY} .   
\end{equation}
After Eqs. (\ref{Sg}) and (\ref{inv}) we can determine the inverted four vectors $\boldsymbol{S} (x,y)$ and the corresponding $\Delta (x,y)$ in Eq. (\ref{tAB}). Finally this leads to the inferred noiseless joint distribution
\begin{equation}
\label{p}
p_{b,R} (x_1,y_1,x_2,y_2 ) = \langle \psi | U^\dagger \Delta_1 (x_1,y_1,) \otimes \Delta_2 (x_2,y_2 ) U |\psi \rangle_b  ,
\end{equation}
so that
\begin{equation}
p_{b,R} (x_1,y_1,x_2,y_2) = \frac{1}{16} \left [ 1 +  \boldsymbol{S} (x_1,y_1) R \boldsymbol{S}^\ast_b (x_2,y_2) \right ]  .
\label{ijd}
\end{equation}

\bigskip

We have shown in Ref. \cite{MAL20} that violation of Bell's inequalities is equivalent to a pathological distribution $p_{b,R} (x_1,y_1,x_2,y_2)$ in the sense of taking negative values. These negative values are actually a clear nonclassical signature since they can never occur in the classical domain where the inversion procedure always leads to the corresponding exact joint probability distribution \cite{AL16a,AL16b,LM17}. Moreover these negativities comply with Fine's theorem \cite{AF82}. This implies that  $ \Delta \left ( x, y \right )$ are no longer positive definite. We can consider this as a signature of $\Delta \left ( x, y \right )$ being a nonclassical detector as a necessary condition to obtain statistics beyond the reach of classical physics \cite{LA17}. 

\bigskip

In the spirit of this work, the violation of the Bell inequalities becomes then a convenient form of disclosing the nonclassical nature of  $\Delta \left ( x, y \right )$ via the  self-tomography protocol developed above. To illustrate this point let us assume that $\boldsymbol{S}_X$, $\boldsymbol{S}_Y$, and $\boldsymbol{S}_{XY}$ are mutually orthogonal, i. e.,
\begin{equation}
   \boldsymbol{S}_X = (1,0,0), \quad \boldsymbol{S}_Y = (0,1,0), \quad \boldsymbol{S}_{XY} = (0,0,1), 
\end{equation}
along with $R = R_0$ and $b=z$, so that the inferred joint distribution (\ref{ijd}) becomes 
\begin{equation}
p_{z,0} (x_1,y_1,x_2,y_2) = \frac{1}{16} \left [ 1 +  \boldsymbol{S} (x_1,y_1) \cdot  \boldsymbol{S}^\ast_z (x_2,y_2) \right ]  .
\end{equation}
being
\begin{equation}
\boldsymbol{S} (x_1,y_1 ) \cdot  \boldsymbol{S}^\ast_z (x_2,y_2) = x_1 x_2 - y_1 y_2v + x_1 x_2 y_1 y_2 \frac{\gamma^2_{XY}}{\gamma^2_X \gamma^2_Y}. 
\end{equation}
Clearly $p_{z,0} (x,y,-x,y) <0$ since 
\begin{equation}
  p_{z,0} (x,y,-x,y) = - \frac{1}{16} \left ( 1 + \frac{\gamma^2_{XY}}{\gamma^2_X \gamma^2_Y} \right ) <0,
\end{equation}
revealing the quantum nature of the detectors. 

\bigskip

We have presented a rather intuitive model of self-tomography. This exploits a rather unique feature of quantum mechanics where detectors are characterised by field states, establishing a fruitful symmetry between observing an observed systems. 

Standard detector tomography may be performed by sending classical-like states to a single realization of the detector. In this regard, the self-tomographic scheme proposed in this work does not pretend to offer practical or technical advantages, but instead focus on conceptual issues. Our main aim here is the possibility to directly check the quantum nature of detection processes, without relying on previous assumptions about the nature of field states. 

\section*{Funding}
Spanish Ministerio de Econom\'ia y Competitividad Project No. FIS2016-75199-P.


\section*{Disclosures}
The authors declare no conflicts of interest.


\begin{thebibliography}{00}

\bibitem{LA17}
A. Luis and L. Ares, "Apparatus contribution to observed nonclassicality," Phys. Rev. A {\bf 102}, 022222 (2020).

\bibitem{LS99a}
A. Luis and L. L. S\'anchez-Soto, "Complete Characterization of Arbitrary Quantum Measurement Processes," Phys. Rev. Lett. {\bf 83}, 3573--3576  (1999).

\bibitem{LS99b}
J. S. Lundeen, A. Feito, H. Coldenstrodt-Ronge, K. L. Pregnell, Ch. Silberhorn, T. C. Ralph, J. Eisert, M. B. Plenio, and I. A. Walmsley, "Tomography of quantum detectors," Nature Phys. {\bf 5}, 27--30 (2009).

\bibitem{LS99c}
L. Zhang, H. B. Coldenstrodt-Ronge, A. Datta, G. Puentes, J. S. Lundeen, X.-M. Jin, B. J. Smith, M. B. Plenio, and I. A. Walmsley, "Mapping coherence in measurement via full quantum tomography of a hybrid optical detector," Nature Photon. {\bf 6},  364--368 (2012).

\bibitem{JF01}
J. Fiurasek, "Maximum-likelihood estimation of quantum measurement," Phys. Rev. A {\bf 64}, 024102 (2001).

\bibitem{ZDCJEPW12}
L. Zhang, A. Datta, H. B. Coldenstrodt-Ronge, X.-M. Jin, J. Eisert, M. B. Plenio, and I. A. Walmsley, "Recursive quantum detector tomography," New. J. Phys. {\bf 14}, 115005 (2012). 

\bibitem{BKSSV17}
M. Bohmann, R. Kruse, J. Sperling, C. Silberhorn, and W. Vogel, "Direct calibration of click-counting detectors," Phys. Rev. A {\bf 95}, 033806 (2017).

\bibitem{NJ19}
R. Nehra, K. V. Jacob, "Characterizing quantum detectors by Wigner functions,"
arXiv:1909.10628 [quant-ph].

\bibitem{DML04}
G. M. D'Ariano, L. Maccone, and P. Lo Presti, "Quantum calibration of measurement instrumentation," Phys. Rev. Lett. {\bf 93}, 250407 (2004).

\bibitem{ZZZ20}
A. Zhang, J. Xie, H. Xu, K. Zheng, H. Zhang, Y.-T. Poon, V. Vedral, and L. Zhang, "Experimental Self-Characterization of Quantum Measurements," Phys. Rev. Lett. {\bf 124}, 040402 (2020).

\bibitem{LS98}
A. Luis and L. L. S\'anchez-Soto, "Conditional generation of field states in parametric down-conversion," Phys. Lett.  A {\bf 244}, 211-216 (1998). 

\bibitem{SVA12}
J. Sperling, W. Vogel, and G. S. Agarwal, "True photocounting statistics of multiple on-off detectors," Phys. Rev. A {\bf 85}, 023820 (2012).

\bibitem{JB19}
M. J\"{o}nsson and Gunnar Bj\"{o}rk, " Evaluating the performance of photon-number-resolving detectors," Phys. Rev. A {\bf 99}, 043822 (2019).

\bibitem{PLFM20}
J. Provazn\'{i}k, L. Lachman, R. Filip, and P. Marek, "Benchmarking photon number resolving detectors," Opt. Express {\bf 28}, 14839 (2020).

\bibitem{HA14}
J.J. Halliwell, Two proofs of Fine's theorem, Phys. Lett. A {\bf 378}, 2945--2950 (2014).

\bibitem{MAL20}	
E. Masa, L. Ares, and A. Luis, "Nonclassical joint distributions and Bell measurements,"  Phys. Lett. A {\bf 384}, 126416 (2020).

\bibitem{Muynck}
W. M. Muynck, \textit{Foundations of Quantum Mechanics, an Empiricist Approach},  (Kluwer Academic Publishers,  2002).

\bibitem{AL16a}
A. Luis, "Nonclassical states from the joint statistics of simultaneous measurements," arXiv:1506.07680 [quant-ph]

\bibitem{AL16b}
A. Luis, "Nonclassical light revealed by the joint statistics of simultaneous measurements," Opt. Lett. {\bf 41}, 1789-1792 (2016).

\bibitem{LM17}
A. Luis and L. Monroy, "Nonclassicality of coherent states: Entanglement of joint statistics," Phys. Rev A {\bf 96}, 063802 (2017).

\bibitem{AF82}
A. Fine, Hidden Variables, "Joint Probability, and the Bell Inequalities," Phys. Rev. Lett {\bf 48}, 291-295 (1982).

\end{thebibliography}
\end{document}